\documentclass[aps,prl,showpacs,amsmath,amssymb,amsfonts,twocolumn,
superscriptaddress,floatfix,footinbib]{revtex4-1}

\usepackage{subfigure}
\usepackage{graphicx}
\usepackage[flushmargin]{footmisc}
\usepackage{color}
\usepackage{soul}
\usepackage{longtable}
\usepackage[normalem]{ulem}
\usepackage{amsmath}
\usepackage{hyperref}

\begin{document}

\title{Propagation and amplification dynamics of 1D polariton condensates}

\author{E. Wertz}
\affiliation{CNRS-Laboratoire de Photonique et Nanostructures, Route de Nozay, 91460 Marcoussis, France}
\author{A. Amo}
\affiliation{CNRS-Laboratoire de Photonique et Nanostructures, Route de Nozay, 91460 Marcoussis, France}
\author{D. D. Solnyshkov}
\affiliation{Institut Pascal, PHOTON-N2, Clermont Université, University Blaise Pascal, CNRS, 24 avenue des Landais, 63177 Aubi\`{e}re cedex, France}
\author{L. Ferrier}
\affiliation{CNRS-Laboratoire de Photonique et Nanostructures, Route de Nozay, 91460 Marcoussis, France}
\author{T.~C.~H.~Liew}
\affiliation{Mediterranean Institute of Fundamental Physics, 31, via Appia Nuova, 00040, Rome, Italy}
%\altaffiliation[Present address: ]{School of Physical and Mathematical Sciences, Nanyang Technological University, 637371, Singapore}
\author{D.~Sanvitto}
%\altaffiliation[Present address: ]{Istituto Nanoscienze � CNR Via Arnesano, 73100 Lecce, Italy}
\affiliation{NNL, Istituto Nanoscienze - CNR, Via Arnesano, 73100 Lecce, Italy}
\affiliation{Istituto Italiano di Tecnologia, Via Barsanti, 73010 Lecce, Italy}
\author{P.~Senellart}
\affiliation{CNRS-Laboratoire de Photonique et Nanostructures, Route de Nozay, 91460 Marcoussis, France}
\author{I.~Sagnes}
\affiliation{CNRS-Laboratoire de Photonique et Nanostructures, Route de Nozay, 91460 Marcoussis, France}
\author{A.~Lema\^{i}tre}
\affiliation{CNRS-Laboratoire de Photonique et Nanostructures, Route de Nozay, 91460 Marcoussis, France}
\author{A. V. Kavokin}
\affiliation{Physics and Astronomy School, University of Southampton, Highfield, Southampton, SO171BJ, UK}
\affiliation{Laboratoire Charles Coulomb, CNRS-Universite de Montpellier II, Pl. Eugene de Bataillon, Montpellier, 34095, cedex, France}

\author{G. Malpuech}
\affiliation{Institut Pascal, PHOTON-N2, Clermont Université, University Blaise Pascal, CNRS, 24 avenue des Landais, 63177 Aubi\`{e}re cedex, France}
\author{J. Bloch}
\email[]{jacqueline.bloch@lpn.cnrs.fr}
\affiliation{CNRS-Laboratoire de Photonique et Nanostructures, Route de Nozay, 91460 Marcoussis, France}

\date{\today}

\begin{abstract}
The dynamics of propagating polariton condensates in one-dimensional microcavities is investigated through time resolved experiments. We find a strong increase in the condensate intensity when it travels through the non-resonantly excited area. This amplification is shown to come from bosonic stimulated relaxation of reservoir excitons into the polariton condensate, allowing for the repopulation of the condensate through non-resonant pumping. Thus, we experimentally demonstrate a polariton amplifier with a large band width, opening the way towards the transport of polaritons with high densities over macroscopic distances.
\end{abstract}

% insert suggested PACS numbers in braces on next line
\pacs{71.36.+c, 67.85.Hj, 78.55.Cr,  78.67.Pt}
% insert suggested keywords - APS authors don't need to do this
%\keywords{}

%\maketitle must follow title, authors, abstract, \pacs, and \keywords
\maketitle

Optical amplifiers are devices with an active media capable of directly amplifying an optical signal traversing them. The medium is non-resonantly pumped, and amplification of a coherent propagating beam is obtained in a single passage by bosonic stimulated emission~\cite{Desurvire1987}. Such devices are important for the regeneration of fibred laser signals transmitted over long distances, or for high  power lasers. Contrary to parametric amplifiers relying on four wave mixing, optical amplifiers based on non-resonant pumping have the great advantage of a large bandwidth and absence of phase matching restrictions.

Such a mechanism of optical amplification can in principle be used to regenerate a coherent population in other bosonic systems, for instance, to produce continuous matter-wave lasing in ultracold atomic Bose-Einstein condensates~\cite{Holland1996}. In an atom laser, a trapped Bose-Einstein condensate is coupled to an output mode, in which atoms flow away while keeping the coherence properties of the condensed state~\cite{Ottl2005}.
The duration of the atom laser is given by the size of the initial condensate and the outcoupling rate~\cite{Hagley1999,Bloch1999}, and it could be extended by adding reservoir atoms feeding and amplifying the matter-wave condensate by bosonic stimulated relaxation.
%%%%%%
%%%%%This duration could be extended by adding reservoir atoms, which would populate the lasing mode by bosonic stimulated relaxation. This reservoir would then feed and amplify the matter-wave condensate. 
While some progress has been done in this direction~\cite{Robins2008}, the continuous refilling of the ultracold atomic cloud remains experimentally challenging and, up to now, atomic lasers can only work in a pseudo-continuous mode with a finite duration.

\begin{figure*}[t]
\includegraphics[width= 1\textwidth]{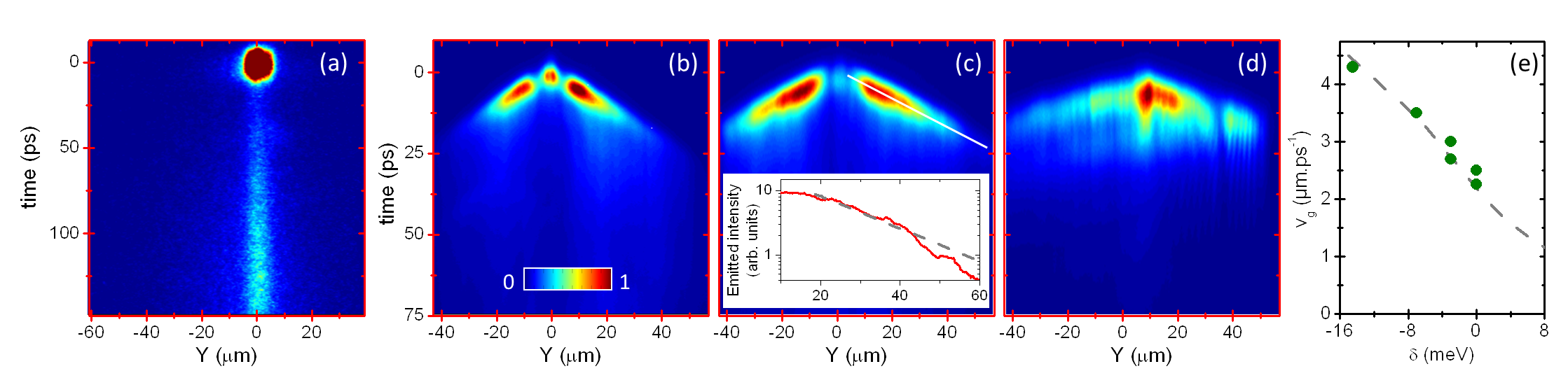}
\caption{(Color online) Time resolved spatial distribution of the polariton emission below condensation threshold at $\delta$ = $-3$ meV (a), and above condensation threshold at $\delta$ = $+3$ meV (b), $\delta$ = $0$ meV (c), and $\delta$ = $-10$ meV (d). In (b)-(d), the excitation density is chosen to produce ejected polariton condensates with a kinetic energy of 1.5~meV. The inset in (c) shows the decay of the polariton density along the white line (the dashed line is an exponential fit). (e) Measured polariton group velocity for different detunings (the dashed line is a guide to the eye). %$L_{x}$ = 3.5 $\mu$m
}\label{Fig1}
\end{figure*}

Similar strategies can be considered for condensates of exciton-polaritons in semiconductor microcavities. Polaritons are mixed light-matter quasi-particles behaving as bosons in the low density limit~\cite{Kasprzak2006}. Polariton condensates present a strong interest both for the study of the fundamental properties of quantum gases (superfluidity~\cite{Amo2008}, quantised vortices~\cite{Lagoudakis2008} and dark~\cite{Amo2011} and bright~\cite{Sich2012} soliton formation have been reported), and for their potential functionalities as integrated optical elements. They present the flexibility of out of resonance optical excitation for their generation and, thanks to their interactions with uncondensed excitons, they can be easily accelerated, propagating over macroscopic distances~\cite{Wertz2010,Christmann2012}. Several works have proposed the use of such propagating polariton condensates to realize a polariton Berry phase interferometer~\cite{Shelykh2009} or various innovative spinoptronic devices~\cite{Liew2008,Johne2010,Amo2010,Shelykh2010b, Flayac2011b} taking advantage of their non-linear properties. However, the finite polariton lifetime (a few tens of picoseconds) results in the decay of the population along the propagation. This effect limits the region in which the polariton density is high enough for particle interactions to induce non-linear phenomena. Thus, it is important to be able to amplify polariton condensates during their propagation.

In this letter, we experimentally demonstrate polariton amplification by a reservoir of uncondensed excitons in a 1D microcavity. We monitor polariton propagation in a time resolved experiment. We show that the polariton condensate propagates with a group velocity that can be controlled varying the exciton-photon spectral detuning. Moreover, amplification is evidenced in the spatial region where excitons are non-resonantly injected: strong increase of the emission is observed when polariton condensates traverse this region. This takes place due to bosonic stimulated relaxation of uncondensed polaritons, and it is well reproduced by solving a modified Gross-Pitaevskii equation describing both the polariton condensate and the excitonic reservoir. Such polariton amplification opens the way to the implementation of cascadable polariton functionalities and optical transistor operation under non-resonant pumping. %\sout{These experiments directly reveal the regeneration of the polariton condensate.}
%The amplification mechanism is based on bosonic stimulated relaxation of polaritons, similarly to an optical amplifier. The observed features can be well reproduced solving a modified Gross-Pitaevskii equation describing both the polariton condensate and the excitonic reservoir, and taking into account exciton relaxation into the polariton condensate via bosonic stimulation. Such polariton amplification is one of the key prerequisites for the implementation of cascadable polariton functionalities and optical transistor operation under non-resonant pumping.

The sample used in our experiments is described in detail in Ref.~[\onlinecite{Wertz2010}]. It consists in a high quality factor $\lambda$/2 planar cavity ($Q$ $\sim$ 16000) containing 12 GaAs quantum wells. Wire cavities with a width of 3.5~$\mu$m and a length of 200~$\mu$m were fabricated using electron beam lithography and reactive ion etching. The cavity wedge, present already in the planar structure, allows for the tuning of the bare exciton-photon energy difference $\delta = E_{C}-E_{X}$ of each wire. Within each wire, $\delta$ remains constant, as they are etched in the direction perpendicular to the wedge. Single microwires are excited non-resonantly (typically 100~meV above the lower polariton energy) with a Ti:Sapphire laser delivering 2~ps pulses with a 82~MHz repetition rate. The laser is focused down to a 3~$\mu$m diameter spot with a microscope objective. The sample emission is collected through the same objective and imaged on the entrance slit of a streak camera coupled or not to a spectrometer in order to measure the time- and energy- or space-resolved polariton propagation. Momentum space is accessed by imaging the Fourier plane of the microscope objective taking advantage of the one-to-one relation between angle of emission and in-plane momentum of polaritons. All experiments are performed at 10~K.

Polariton condensation in such microwires has been reported under cw excitation~\cite{Wertz2010} showing the formation of extended polariton condensates propagating out of the excitation region. Here, space- and time-resolved measurements allow direct monitoring of the propagation dynamics, as shown in Fig.~\ref{Fig1}(a)-(d) for %3.5~$\mu$m wide 
microwires with different exciton-cavity detunings under non-resonant pulsed pumping at $Y=0$, in the middle of the wire. Below the condensation threshold [Fig.~\ref{Fig1}(a)] the emission dynamics is slow, reflecting the occupation of the excitonic reservoir~\cite{Bloch1997}, and very little propagation is observed: the emission remains localized in the area beneath the pulsed pump laser. The rise of the emission intensity apparent in the reported time window is given by the dynamics of exciton formation and relaxation into low momentum states~\cite{Damen1990,Szczytko2004}. Nonetheless, the total carrier population continuously decays in time.

At carrier densities above the condensation threshold ($P_{th}$), stimulated polariton relaxation gives rise to the formation of a condensate in the region of excitation at the bottom of the lower polariton branch with zero in-plane momentum ($k=0$). At the high excitation densities reported in Fig.~\ref{Fig1}(b)-(d) ($50~P_{th}$) the condensation takes place few picoseconds after the arrival of the excitation pulse ($t=0$). The presence of an uncondensed excitonic reservoir in the excitation region induces a blueshift of the polariton energy (4~meV) caused by repulsive exciton-polariton interactions~\cite{Wertz2010, Ferrier2011}. Since the excitonic reservoir remains localized in the region of excitation (because of the large exciton effective mass), no blueshift is generated outside this area. In this way, the reservoir induces an interaction energy that can be transformed into kinetic energy when the condensate exits the excitation region, resulting in the acceleration and propagation of the polariton condensate on both sides of this region, as observed in Fig.~\ref{Fig1}(b)-(d). In the high excitation density case reported in Fig.~\ref{Fig1}, part of the condensate energy ($\sim$2.5~meV) is relaxed due to parametric polariton-polariton inetractions, as reported in Ref.~\onlinecite{Tanese2012} (see also~\footnote{See Supplementary Information for a complete description of the model and additional results under high density excitation}). The rest of the interaction energy ($\sim$1.5~meV) is transformed into kinetic energy setting the condensate in movement.

Once the condensate exits the excitation spot, the propagation velocity remains constant and can be directly measured from the slope of the polariton trajectory in the real space-time resolved images~\cite{Freixanet2000}. The group velocity in the case of Fig.~\ref{Fig1}(b) is $1.8~\mu$m$\cdot$ps$^{-1}$, and it is directly given by the kinetic energy. % equal to the polariton blueshift at the excitation region. %It is directly given by the first derivative of the polariton dispersion at the in-plane wavector corresponding to a kinetic energy equal to the polariton blueshift at the excitation region.
The group velocity is inversely proportional to the square root of the polariton mass, which is fixed by the curvature of the polariton dispersion~\cite{Wertz2010}. Thus, by changing this curvature we can control the group velocity of the expelled polariton packets. This is shown in Fig.~\ref{Fig1}(b)-(d) for three different detunings of $+3$, $0$ and $-10$~meV, respectively. In all three cases we selected excitation densities giving rise to the same kinetic energies for the ejected condensates
%reservoir induced blueshift
 (1.5~meV). When the detuning is decreased towards negative values, polaritons get lighter and the propagation speed increases, as summarized in Fig.~\ref{Fig1}(e).

The inset of Fig.~\ref{Fig1}(c) shows the characteristic decay of the density as the polariton wavepackets propagate away from the excitation region (propagation length of $\sim 17~\mu$m). This decay is given by the photon lifetime in our etched microcavities, which is on the order of 15~ps, and evidences the need for an amplification or repumping of the condensate if propagation over longer distances is desired. We propose an amplification mechanism similar to that of optical amplifiers. It is based on the stimulated relaxation of the non-resonantly excited reservoir of excitons/polaritons to the bottom of the lower polariton branch, induced by the presence of a large polariton population in that state. 

\begin{figure}[t]
\includegraphics[width=\columnwidth]{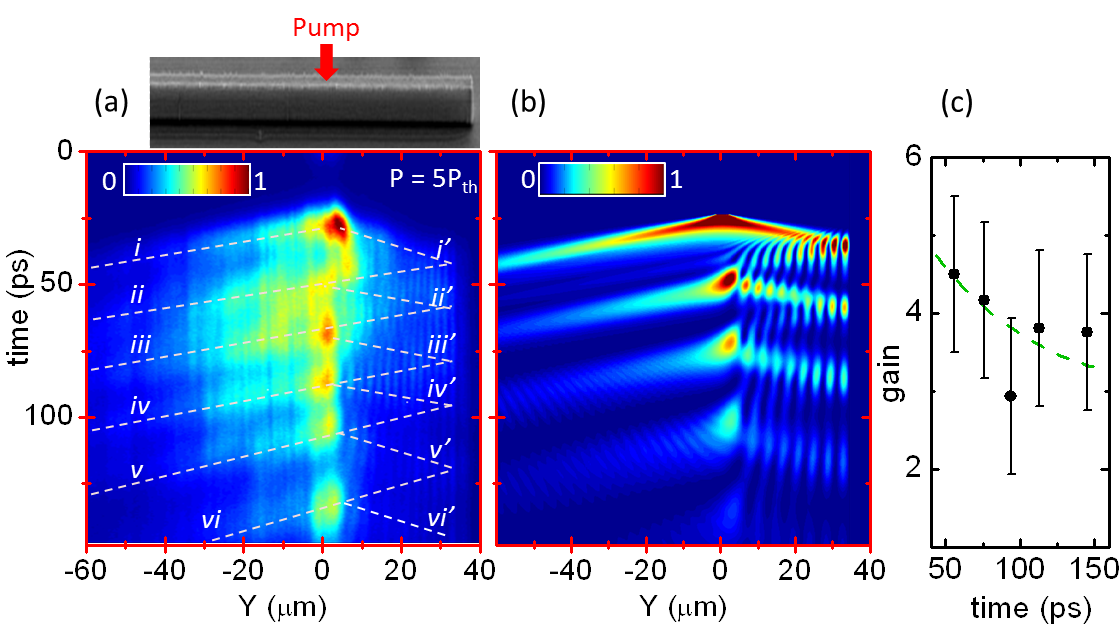}
\caption{(Color online) (a) Time resolved spatial distribution of the polariton emission at $5~P_{th}$ in a wire with $\delta$ = $-3$ meV. The excitation spot is located 35~$\mu$m from the right edge of the wire.%for different pump powers as indicated in the figure; (a-b)
% measured experimentally and 
 (b) Simulation considering polariton condensates propagating and interacting with the excitonic reservoir. (c) Measured amplification gain at each of the successive repopulations of the condensate; the green dotted line is an exponential fit with a decay of 300~ps. The modulation of the intensity observed both in (a) and (b) at Y$>0$ arises from the interference between incoming and reflected condensates.% $L_{x}$ = 3.5 $\mu$m and $\delta$ = $-3$ meV.
 }\label{Fig2}
\end{figure}

In order to experimentally prove polariton amplification, we place the excitation spot 35~$\mu$m away from one of the edges of the wire, as schematically represented in Fig.~\ref{Fig2}(a). % This distance is on the order of the propagation distance.
%Figure~\ref{Fig2}(a) shows the polariton emission below the condensation threshold. The dynamics are very slow reflecting the occupation of the excitonic reservoir~\cite{Bloch1997}, and very little propagation is observed: the emission remains localized in the area beneath the pulsed pump laser. Though a rise of the emission intensity is observed in the reported time window, given by the dynamics of exciton formation and relaxation into low momentum states~\cite{Damen1990,Szczytko2004},the total carrier population continuously decays in time. %The observed increase comes from the dynamics of the exciton formation and relaxation into low momentum states~\cite{Damen1990,Szczytko2004}.
Above threshold [$5~P_{th}$, Fig.~\ref{Fig2}(a)], the condensate is formed at $t_{1}=25$~ps after the arrival of the excitation pulse, and %the behavior reported in Fig.~\ref{Fig1} is recovered, with a condensate 
and two condensates are ejected from the excitation region, one departing to the left (labeled \emph{i}) and another one to the right (labeled \emph{i'}). When condensate \emph{i'} reaches the edge of the wire, it is reflected and propagates back. Remarkably, when it gets back to the pumped area (at $t_{2}=50$~ps) a strong increase of the emission signal is observed, accompanied by the propagation of two new condensate packets (\emph{ii} and \emph{ii'}), to the right and to the left of the region where the reservoir is located. The regeneration of the condensate takes place five times in our observation window. To evaluate the gain of the condensate amplification we first measure the intensity of the signal close to the edge of the wire (Y=32~$\mu$m), and then we estimate from this value what the signal intensity would be without amplification at Y=-15~$\mu$m (considering only the polariton decay). Finally we take the ratio of this estimated intensity to the measured one at this very same position. The results are reported on Fig.~\ref{Fig2}(c), showing a gain of more than 4, and a time decay of the gain on the order of 300~ps (fitted dashed lines), compatible with that of the reservoir.

\begin{figure}[t]
\includegraphics[width= 0.7\columnwidth]{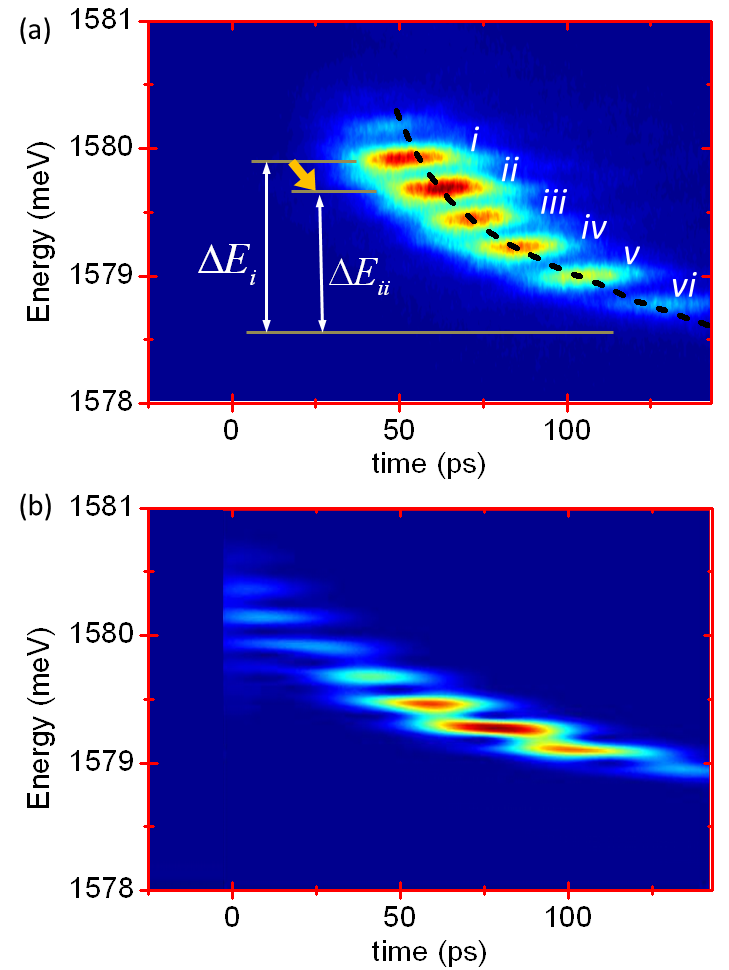}
\caption{(Color online) Energy of the successive condensates as they pass through Y=-25~$\mu$m on Fig.~\ref{Fig2}(b): (a) measured experimentally and (b) simulated. The black dashed line is the measured energy at $k=0$ under the excitation spot.}\label{Fig3}
\end{figure}

In order to understand in detail the amplification dynamics of the polariton signal, we study the reservoir induced blueshift and the emission energy of the subsequently expelled condensates. This is experimentally explored in Fig.~\ref{Fig3}(a), which shows the energy of the condensates traversing the region at Y=-25~$\mu$m on Fig.~\ref{Fig2}(a). At $t_{1}=25$~ps condensation takes place at $k=0$ in the pumped region with an energy that is blueshifted by an amount $\Delta E_{i}=1.5$~meV with respect to the regions without reservoir [see Fig.~\ref{Fig3}(a)].
Note that in this case of moderate excitation density (5~$P_{th}$) polariton-parametric relaxation is negligible (contrary to the case of Fig.~\ref{Fig1}). $\Delta E_{i}$ corresponds to the condensate energy in the excitation region, given by the total reservoir population at $t_{1}$, and sets the kinetic energy of condensates \emph{i} and \emph{i'}~\cite{Wertz2010}. The reservoir induced interaction energy is thus fully converted into kinetic energy. At $t_{2}$, the reservoir population has partly decayed resulting in a decrease of the $k=0$ energy in the region of excitation ($\Delta E_{ii}=1.3$~meV). The arrival in that region of the reflected condensate \emph{i'} induces a re-stimulation of the condensate from the reservoir, %mostly by polariton-polariton interactions,
 amplifying the condensate density. This re-stimulation takes place simultaneously with the energy relaxation of condensate via polariton-reservoir pair scattering, down to the local potential energy set by the reservoir density at the time delay $t_{2}$ (orange arrow in Fig.~\ref{Fig3}).

% the stimulated relaxation of polaritons from the reservoir, thus amplifying the condensate density. However, stimulated relaxation takes place, not to the energy of the condensate \emph{i'} but to the lowest energy state ($k=0$) in the reservoir region at the time delay $t_{2}$. This state is now at a lower energy than that of condensate \emph{i'} because at $t_{2}$ the total number of reservoir excitons is smaller, and so is the blueshift $\Delta E_{ii}$ ($<\Delta E_{i}$). Stimulated relaxation gives rise to two new condensates \emph{ii} and \emph{ii'} at $t_{2}$, that are ejected with an energy given by $\Delta E_{ii}$.%slightly smaller than that of condensates \emph{i} and \emph{i'}.

%Figure~\ref{Fig3}(a) shows the energy of the cascade of condensates \emph{i}, \emph{ii}, \emph{iii}, etc. when they pass the region around Y=-25~$\mu$m on Fig.~\ref{Fig2}(b). 

The mechanism we have described presents strong similarities to that of an optical amplifier, in which bosonic stimulation takes place exactly to the same energy state as that of the incoming field, which acts a  seed. In our case, stimulation takes place simultaneously with the inelastic relaxation of the incoming condensate. The cascade observed in Fig.~\ref{Fig3} reflects the time evolution of the $k=0$ energy in the excitation region [measured in dashed lines in Fig.~\ref{Fig3}(a)], sampled by the passage of the successive condensates.
%The energy of the condensate exhibits step-like behavior (\emph{i'}, \emph{ii'}. \emph{iii'}, \ldots) because of the relaxation of the condensate when entering the region with a highly dense reservoir. This relaxation takes place through inelastic scattering via interaction with reservoir excitons. %From the observed blueshift in the excitation region, we extract an exciton density of up to XX $\mu$m$^{-1}$.
The relaxation of the condensate traversing the reservoir region is, thus, essential to account for this energy cascade. As $\Delta E$ sets the kinetic energy of the expelled condensates, the energy relaxation results in the slow down of the subsequent emitted condensates, as it can be observed from the increasing slope of the dashed lines in Fig.~\ref{Fig2}(a).
We can model the coupled dynamics of the condensate and the reservoir, with the use of a modified Gross-Pitaevskii equation including terms for the re-stimulation and inelastic relaxation of the condensate~\cite{Choi1998,Wouters2010c,*Wouters2010d} depending on the local reservoir density~[25]. This model allows to fully reproduce our results, as evidenced in Figs.~\ref{Fig2}(b) and \ref{Fig3}(b), including the slow down of the successively amplified condensates.

\begin{figure}[t]
\includegraphics[width= \columnwidth]{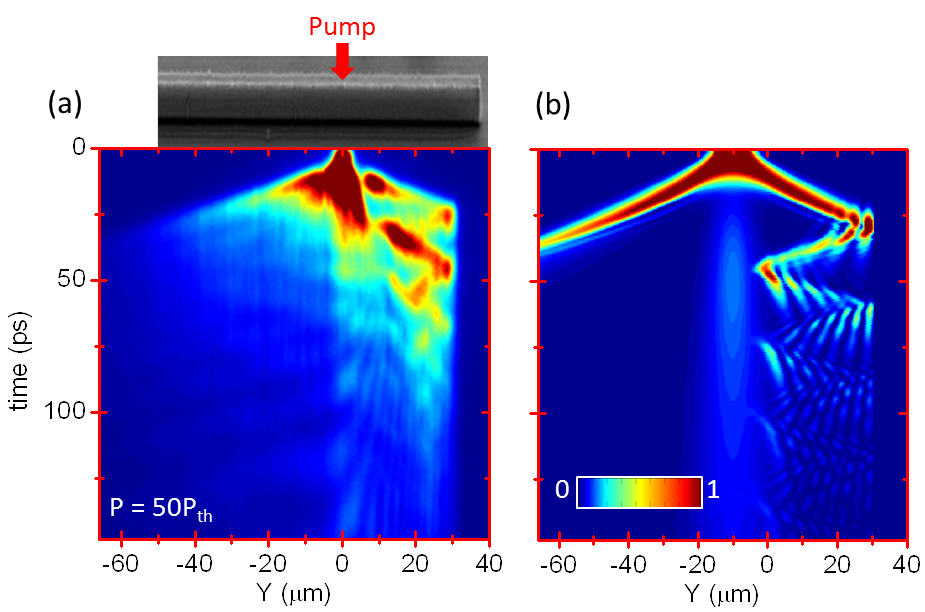}
\caption{(Color online) Time resolved spatial distribution of the polariton emission at very high pump powers (50 times the condensation threshold) for a wire with $\delta$ = $-3$ meV. The reservoir creates a potential barrier that cannot be crossed by the condensate: (a) measured experimentally and (b) simulated.% $L_{x}$ = 3.5 $\mu$m and $\delta$ = $-3$ meV.
}\label{Fig4}
\end{figure}

The flexibility of the out-of resonant excitation provides supplementary control of the propagating condensates. If the initial injected density is %very high, it was shown in Ref.~\cite{Tanese2012} that interactions are responsible for the generation of polariton condensates energies lower than the blueshift 
high enough, polariton-polariton interactions close to the
excitation area induce polariton scattering towards lower-
energy states~\cite{Tanese2012}, as discussed for the case of Fig.~\ref{Fig1}. % created by the reservoir in the excitation region.
%In this regime, the reservoir creates a potential barrier of higher energy than the initially ejected condensates.
The situation with excitation close to the wire edge is shown in Fig.~\ref{Fig4}(a), where %at short time delays, one condensate is ejected on each side of the excitation spot, with energies smaller than those of the reservoir-induced barrier~\cite{Wertz2010,Tanese2012}.
the condensate that is initially ejected towards the right has a smaller energy than the reservoir-induced barrier~\cite{Gao2012} %in Fig.~\ref{Fig4}(a)
and gets trapped between the excitation region and the edge of the wire~\cite{Wertz2010,Tosi2012}, %and the barrier, and 
rebounding continuously between the two without appreciable energy loss~[25]. The spatial oscillation of this polariton wave-packet is well reproduced by our simulations when run with a high initial density [Fig.~\ref{Fig4}(b)].

We have evidenced the amplification of polariton condensates under non-resonant excitation. The propagation with controlled velocity along with the re-population process shown here, are two key ingredients to achieve transport of polariton condensates over macroscopic distances with a high density. Our scheme provides a large energy bandwidth for the re-populating beam, ranging from the gap energy of the employed quantum wells to the absorption energy of the materials constituting the Bragg mirrors of the cavity (from $\sim$1.61~eV to 1.83~eV in our case). These are pre-requisites for the implementation of a number of proposed functionalities based on the propagation of coherent polariton condensates~\cite{Liew2008, Shelykh2009, Johne2010, Amo2010, Flayac2011b,Ballarini2012}. In particular, our results open the way to their inter-connection and cascadability, with the great advantage of being an out-of-resonance scheme not requiring the phase matching conditions of configurations based on parametric scattering~\cite{Amo2009,Adrados2011,Sich2012}.

This work was partly supported by the \emph{RTRA Triangle de la Physique} "Boseflow1D", by the contract ANR-11-BS10-001 "Quandyde", by the FP7 \emph{ITNs} "Clermont4" (235114) and  "Spin-Optronics" (237252). T.L. was supported by the FP7 Marie Curie project "EPOQUES" (298811).

\newpage
\section{Supplementary Information}
\subsection{Condensation and relaxation model}

In this supplementary, we describe the theoretical model used to simulate polariton propagation, amplification, and relaxation in 1D wire microcavities. As a main tool, we have used the Gross-Pitaevskii equation for polaritons in the parabolic approximation valid at low wavevectors, which is the regime observed in the experiments. This Gross-Pitaevskii equation has been extended to take into account the stimulated replenishing of the propagating condensate and its energy relaxation due to the interaction of polaritons with the reservoir of uncondensed excitons. The latter are strongest when the condensate crosses the region of space where the reservoir is present, localized under the excitation spot. The relaxation mechanism assumes a thermalized reservoir in which excitonic lower energy states are more populated than higher energy ones. In that case, dissipation of the condensate energy takes place because the probability for condensate polaritons to scatter down is larger than to scatter up, since it is easier to find a low-energy exciton in the reservoir which would go up (conserving the total energy), than to find a high energy one able to go down. Once the polariton-reservoir exciton scattering has occurred, the extra energy gained by the reservoir is quickly dissipated by phonons, which interact strongly with excitons. For the condensed polaritons, interactions with phonons have been shown to play a negligible role in their relaxation (Ref. 12 of the main text).

The 1D Gross-Pitaevskii equation for polaritons with phenomenologically included lifetime reads:
\begin{widetext}
\begin{equation}
i\hbar \frac{{\partial \psi \left( {x,t} \right)}}{{\partial t}} =  - \frac{{{\hbar ^2}}}{{2{m_{pol}}}}\Delta \psi \left( {x,t} \right)
 + \alpha \left( {{{\left| {\psi \left( {x,t} \right)} \right|}^2} + {n_R}\left( {x,t} \right)} \right)\psi \left( {x,t} \right) - \frac{{i\hbar }}{{2\tau_{pol} }}\psi \left( {x,t} \right)
\label{eq:GPE}
\end{equation}
\end{widetext}
where $m_{pol}=4\times 10^{-5}m_{0}$ is the polariton mass ($m_{0}$ is the free electron mass	), $\alpha=3 E_b a_B^2 /W$ is the polariton-polariton interaction constant ($E_b=6$ meV is the exciton binding energy, $a_B=10~$nm is the exciton Bohr radius and $W=3.5~\mu$m is the width of the wire), $\tau_{pol}=30$~ps is the polariton lifetime.

The stimulation of reservoir excitons towards the condensate 
can be phenomenologically described by a term proportional to the condensate wavefunction and to the reservoir density $n_R$, whose profile is determined by that of the pump:

\begin{equation} + i\gamma {n_R}\left( {x,t} \right)\psi \left( {x,t} \right)
\label{eq:stimulation}
\end{equation}

%The relaxation of polaritons in the condensate is provided mostly by the polariton-\emph{reservoir} interactions, while the phonons have been shown to play a negligible role for a propagating condensate (Ref. 12 of the main text). Moreover, t
The energy towards which the condensate relaxes is determined by the local blueshift created by the reservoir. While crossing the reservoir zone, the polariton packet relaxes towards the top of the potential created by the reservoir, which is exponentially decaying with time. The relaxation of the condensate can be described in the lowest-order approximation by adding a small imaginary part to the Hamiltonian~\cite{Choi1998,Wouters2010c,*Wouters2010d}:

\begin{equation}
i\hbar \frac{{\partial \psi\left(x,t\right) }}{{\partial t}} = \left( {1 - i\Lambda\left(x,t\right) } \right)\left(H-\mu\left(x,t\right)\right)\psi\left(x,t\right)
\end{equation}

where $\Lambda\left(x,t\right)\ll 1$ is a small parameter. The difference with the phenomenological lifetime is that not only the wavefunction is multiplied by this coefficient, but also the difference between the Hamiltonian and the chemical potential of the condensate. $\Lambda\left(x,t\right)$ describes therefore not the decay of the particles, but the decay of the excess energy and its convergence towards $\mu\left(x,t)\right)$. In our case, $\Lambda\left(x,t\right)$ is a function of space and time, and of the reservoir density, since the relaxation takes place only in the region of the reservoir. The same applies also to the chemical potential $\mu\left(x,t\right)=\alpha n_R\left(x,t\right)$ (interactions in the condensate can be neglected in this regime with respect to the potential created by the reservoir \cite{Ferrier2011}). The final Gross-Pitaevskii equation for the condensate dynamics reads:
\begin{widetext}
\begin{equation}i\hbar \frac{{\partial \psi }}{{\partial t}} =  - \frac{{{\hbar ^2}}}{{2m_{pol}}}\Delta \psi  + i\gamma {n_R(x,t)}\psi  - \frac{{i\hbar }}{{2{\tau _{pol}}}}\psi  + \alpha \left( {{{\left| \psi  \right|}^2} + {n_R(x,t)}} \right)\psi  + i\Lambda_0 \frac{{{n_R(x,t)}}}{{\max \left( {{n_R(x,t)}} \right)}}\frac{{{\hbar ^2}}}{{2m_{pol}}}\Delta \psi
\label{eq:full}
\end{equation}
\end{widetext}

\noindent where the relaxation parameter $\Lambda\left(x,t\right)=\Lambda_0 n_R\left(x,t\right)/\max \left( n_R\left(x,t\right)\right)$ contains the reservoir density explicitly (the $\max \left( n_R\left(x,t\right)\right)$ normalizes the relaxation to the maximal density of the reservoir
 at $t=0$. In our simulations, we have taken the fitting parameter $\Lambda_0=0.2$. It is clearly seen that the kinetic energy of the condensate decays because of the relaxation governed by the last term in Eq.~\ref{eq:full}, while the potential energy resulting from the condensate interactions (previous to the last term) does not decay.

Finally, the reservoir dynamics is described by the equation
\[\frac{{\partial {n_R}}}{{\partial t}} = D\Delta {n_R} - \frac{{{n_R}}}{{{\tau _R}}} - \gamma {n_R}{\left| \psi  \right|^2} + P\left( t \right)\exp \left( { - \frac{{{{\left( {x - {x_0}} \right)}^2}}}{{{\sigma ^2}}}} \right)\]
where $D=5\times 10^{-2}$~$\mu$m$^{2}$ ps$^{-1}$ is the diffusion coefficient (fitting parameter obtained from experimental image at low pumping), $\tau_R=300$~ps is the reservoir lifetime, $P=P_{0}\delta(t)$ is the pumping, which directly creates the reservoir population at $t=0$ with $P_{0}=1200 \mu$m$^{-1}$, and $\sigma=4~\mu$m is the width of the pump. We complete the initial conditions by imposing the presence of a condensate at $t=0$ possessing the same spatial profile as the reservoir: $\psi\left(x,0\right)=1\times\exp(-\left(x-x_0\right)^2/\sigma^2)$. 
This initial condition in our simulations accounts for the initial relaxation of polaritons forming the condensate state. Once it is formed, the dominant mechanism populating the condensate is the stimulated scattering described by Eq.~\ref{eq:stimulation}, in which $\gamma$ is a fitting parameter with a value of $3\times10^9\mu$m s$^{-1}$.

The configuration of the present experiment does not allow to apply the self-consistent mixed Boltzmann-Gross-Pitaevskii model for polariton relaxation with renormalization of the condensed states developed recently for microcavity micropillars~\cite{Galbiati2012}, because in the present case the condensation occurs on free propagating states, which are not confined. This is why we used the simplified model corresponding to $T=0$~K, with first order phenomenological relaxation terms. The importance of both re-stimulation and relaxation is clear from Figs. 2 and 3 of the main text. Polariton-polariton scattering within the condensate is fully taken into account by the Gross-Pitaevskii Eqs.~\ref{eq:GPE} or \ref{eq:full}.

\subsection{Time evolution of the condensate energy at high excitation density}

\begin{figure}[t]
\includegraphics[width=1 \columnwidth]{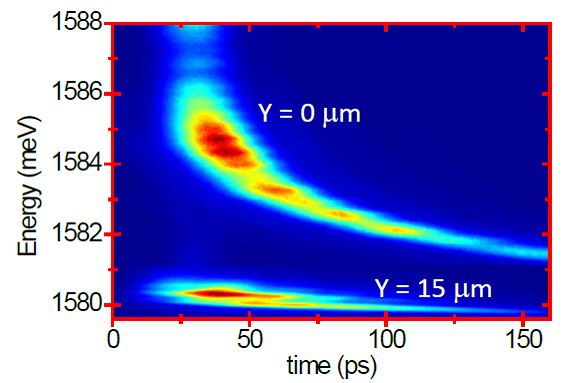}
\caption{(Color online) Experimentally measured time and energy resolved emission in the conditions of Fig.~4 of the main text. The upper trace shows the emission from the region at Y=0, where the excitation spot is located. The lower trace shows the energy measured at Y=15~$\mu$m, where the condensate is trapped and rebounds elastically between the edge of the microwire and the potential barrier created by the reservoir.}

\label{FigS1}
\end{figure}

Figure~S1 shows the measured time and energy resolved emission corresponding to the high excitation density case (50~$P_{th}$) shown in Fig.~4(a). The upper trace shows the energy measured at the position of the excitation spot, coming from the emission from the $k=0$ polaritons that condense in that region. The initial blueshift arises from the strong exciton-polariton interactions, with a decay in time corresponding to the lifetime of the excitonic reservoir. The lower trace shows the emission from a region to the right of the excitation spot, between this and the edge of the microwire. In this case of very high excitation density and large reservoir induced blueshift in the excitation region ($\sim$4~meV), the ejected condensate relaxes towards lower energies via parametric polariton-polariton interactions. This process is more efficient the closer the blueshifted energy is to the inflexion point of the unperturbed polariton dispersion~\cite{Wertz2010,Tanese2012}. For this reason, the initial energy of the ejected condensate is lower than that of the excitation region and only part of the interaction energy is transformed into kinetic energy [a similar situation takes place in Fig.~1(b)-(d) in the main text]. After the ejection, the energy of condensate remains constant, showing that the condensate rebounds elastically between the edge of the wire and the potential barrier created by the reservoir excitons localised under the excitation region.

This regime is very different to that shown in Fig.~2(a) at lower excitation density. In that case the reservoir induced blueshift is smaller (1.5~meV) and  far from the energy corresponding to the inflexion point. Parametric processes are thus not favored and the condensate does not relax to lower energies during the ejection, and the reservoir interaction energy is fully converted into the kinetic energy of the ejected condensate, as discussed in the main text.


\begin{thebibliography}{34}%
\makeatletter
\providecommand \@ifxundefined [1]{%
 \@ifx{#1\undefined}
}%
\providecommand \@ifnum [1]{%
 \ifnum #1\expandafter \@firstoftwo
 \else \expandafter \@secondoftwo
 \fi
}%
\providecommand \@ifx [1]{%
 \ifx #1\expandafter \@firstoftwo
 \else \expandafter \@secondoftwo
 \fi
}%
\providecommand \natexlab [1]{#1}%
\providecommand \enquote  [1]{``#1''}%
\providecommand \bibnamefont  [1]{#1}%
\providecommand \bibfnamefont [1]{#1}%
\providecommand \citenamefont [1]{#1}%
\providecommand \href@noop [0]{\@secondoftwo}%
\providecommand \href [0]{\begingroup \@sanitize@url \@href}%
\providecommand \@href[1]{\@@startlink{#1}\@@href}%
\providecommand \@@href[1]{\endgroup#1\@@endlink}%
\providecommand \@sanitize@url [0]{\catcode `\\12\catcode `\$12\catcode
  `\&12\catcode `\#12\catcode `\^12\catcode `\_12\catcode `\%12\relax}%
\providecommand \@@startlink[1]{}%
\providecommand \@@endlink[0]{}%
\providecommand \url  [0]{\begingroup\@sanitize@url \@url }%
\providecommand \@url [1]{\endgroup\@href {#1}{\urlprefix }}%
\providecommand \urlprefix  [0]{URL }%
\providecommand \Eprint [0]{\href }%
\providecommand \doibase [0]{http://dx.doi.org/}%
\providecommand \selectlanguage [0]{\@gobble}%
\providecommand \bibinfo  [0]{\@secondoftwo}%
\providecommand \bibfield  [0]{\@secondoftwo}%
\providecommand \translation [1]{[#1]}%
\providecommand \BibitemOpen [0]{}%
\providecommand \bibitemStop [0]{}%
\providecommand \bibitemNoStop [0]{.\EOS\space}%
\providecommand \EOS [0]{\spacefactor3000\relax}%
\providecommand \BibitemShut  [1]{\csname bibitem#1\endcsname}%
\let\auto@bib@innerbib\@empty
%</preamble>
\bibitem [{\citenamefont {Desurvire}\ \emph {et~al.}(1987)\citenamefont
  {Desurvire}, \citenamefont {Simpson},\ and\ \citenamefont
  {Becker}}]{Desurvire1987}%
  \BibitemOpen
  \bibfield  {author} {\bibinfo {author} {\bibfnamefont {E.}~\bibnamefont
  {Desurvire}}, \bibinfo {author} {\bibfnamefont {J.~R.}\ \bibnamefont
  {Simpson}}, \ and\ \bibinfo {author} {\bibfnamefont {P.~C.}\ \bibnamefont
  {Becker}},\ }\href@noop {} {\bibfield  {journal} {\bibinfo  {journal} {Opt.
  Lett.}\ }\textbf {\bibinfo {volume} {12}},\ \bibinfo {pages} {888} (\bibinfo
  {year} {1987})}\BibitemShut {NoStop}%
\bibitem [{\citenamefont {Holland}\ \emph {et~al.}(1996)\citenamefont
  {Holland}, \citenamefont {Burnett}, \citenamefont {Gardiner}, \citenamefont
  {Cirac},\ and\ \citenamefont {Zoller}}]{Holland1996}%
  \BibitemOpen
  \bibfield  {author} {\bibinfo {author} {\bibfnamefont {M.}~\bibnamefont
  {Holland}}, \bibinfo {author} {\bibfnamefont {K.}~\bibnamefont {Burnett}},
  \bibinfo {author} {\bibfnamefont {C.}~\bibnamefont {Gardiner}}, \bibinfo
  {author} {\bibfnamefont {J.~I.}\ \bibnamefont {Cirac}}, \ and\ \bibinfo
  {author} {\bibfnamefont {P.}~\bibnamefont {Zoller}},\ }\href {\doibase
  10.1103/PhysRevA.54.R1757} {\bibfield  {journal} {\bibinfo  {journal} {Phys.
  Rev. A}\ }\textbf {\bibinfo {volume} {54}},\ \bibinfo {pages} {R1757}
  (\bibinfo {year} {1996})}\BibitemShut {NoStop}%
\bibitem [{\citenamefont {\"Ottl}\ \emph {et~al.}(2005)\citenamefont {\"Ottl},
  \citenamefont {Ritter}, \citenamefont {K\"ohl},\ and\ \citenamefont
  {Esslinger}}]{Ottl2005}%
  \BibitemOpen
  \bibfield  {author} {\bibinfo {author} {\bibfnamefont {A.}~\bibnamefont
  {\"Ottl}}, \bibinfo {author} {\bibfnamefont {S.}~\bibnamefont {Ritter}},
  \bibinfo {author} {\bibfnamefont {M.}~\bibnamefont {K\"ohl}}, \ and\ \bibinfo
  {author} {\bibfnamefont {T.}~\bibnamefont {Esslinger}},\ }\href@noop {}
  {\bibfield  {journal} {\bibinfo  {journal} {Phys. Rev. Lett.}\ }\textbf
  {\bibinfo {volume} {95}},\ \bibinfo {pages} {090404} (\bibinfo {year}
  {2005})}\BibitemShut {NoStop}%
\bibitem [{\citenamefont {Hagley}\ \emph {et~al.}(1999)\citenamefont {Hagley},
  \citenamefont {Deng}, \citenamefont {Kozuma}, \citenamefont {Wen},
  \citenamefont {Helmerson}, \citenamefont {Rolston},\ and\ \citenamefont
  {Phillips}}]{Hagley1999}%
  \BibitemOpen
  \bibfield  {author} {\bibinfo {author} {\bibfnamefont {E.~W.}\ \bibnamefont
  {Hagley}}, \bibinfo {author} {\bibfnamefont {L.}~\bibnamefont {Deng}},
  \bibinfo {author} {\bibfnamefont {M.}~\bibnamefont {Kozuma}}, \bibinfo
  {author} {\bibfnamefont {J.}~\bibnamefont {Wen}}, \bibinfo {author}
  {\bibfnamefont {K.}~\bibnamefont {Helmerson}}, \bibinfo {author}
  {\bibfnamefont {S.~L.}\ \bibnamefont {Rolston}}, \ and\ \bibinfo {author}
  {\bibfnamefont {W.~D.}\ \bibnamefont {Phillips}},\ }\href {\doibase
  10.1126/science.283.5408.1706} {\bibfield  {journal} {\bibinfo  {journal}
  {Science}\ }\textbf {\bibinfo {volume} {283}},\ \bibinfo {pages} {1706}
  (\bibinfo {year} {1999})}\BibitemShut {NoStop}%
\bibitem [{\citenamefont {Bloch}\ \emph {et~al.}(1999)\citenamefont {Bloch},
  \citenamefont {H\"ansch},\ and\ \citenamefont {Esslinger}}]{Bloch1999}%
  \BibitemOpen
  \bibfield  {author} {\bibinfo {author} {\bibfnamefont {I.}~\bibnamefont
  {Bloch}}, \bibinfo {author} {\bibfnamefont {T.~W.}\ \bibnamefont {H\"ansch}},
  \ and\ \bibinfo {author} {\bibfnamefont {T.}~\bibnamefont {Esslinger}},\
  }\href {\doibase 10.1103/PhysRevLett.82.3008} {\bibfield  {journal} {\bibinfo
   {journal} {Phys. Rev. Lett.}\ }\textbf {\bibinfo {volume} {82}},\ \bibinfo
  {pages} {3008} (\bibinfo {year} {1999})}\BibitemShut {NoStop}%
\bibitem [{\citenamefont {Robins}\ \emph {et~al.}(2008)\citenamefont {Robins},
  \citenamefont {Figl}, \citenamefont {Jeppesen}, \citenamefont {Dennis},\ and\
  \citenamefont {Close}}]{Robins2008}%
  \BibitemOpen
  \bibfield  {author} {\bibinfo {author} {\bibfnamefont {N.~P.}\ \bibnamefont
  {Robins}}, \bibinfo {author} {\bibfnamefont {C.}~\bibnamefont {Figl}},
  \bibinfo {author} {\bibfnamefont {M.}~\bibnamefont {Jeppesen}}, \bibinfo
  {author} {\bibfnamefont {G.~R.}\ \bibnamefont {Dennis}}, \ and\ \bibinfo
  {author} {\bibfnamefont {J.~D.}\ \bibnamefont {Close}},\ }\href@noop {}
  {\bibfield  {journal} {\bibinfo  {journal} {Nature Phys.}\ }\textbf {\bibinfo
  {volume} {4}},\ \bibinfo {pages} {731} (\bibinfo {year} {2008})}\BibitemShut
  {NoStop}%
\bibitem [{\citenamefont {Kasprzak}\ \emph {et~al.}(2006)\citenamefont
  {Kasprzak}, \citenamefont {Richard}, \citenamefont {Kundermann},
  \citenamefont {Baas}, \citenamefont {Jeambrun}, \citenamefont {Keeling},
  \citenamefont {Marchetti}, \citenamefont {Szymanska}, \citenamefont {Andre},
  \citenamefont {Staehli}, \citenamefont {Savona}, \citenamefont {Littlewood},
  \citenamefont {Deveaud},\ and\ \citenamefont {Dang}}]{Kasprzak2006}%
  \BibitemOpen
  \bibfield  {author} {\bibinfo {author} {\bibfnamefont {J.}~\bibnamefont
  {Kasprzak}}, \bibinfo {author} {\bibfnamefont {M.}~\bibnamefont {Richard}},
  \bibinfo {author} {\bibfnamefont {S.}~\bibnamefont {Kundermann}}, \bibinfo
  {author} {\bibfnamefont {A.}~\bibnamefont {Baas}}, \bibinfo {author}
  {\bibfnamefont {P.}~\bibnamefont {Jeambrun}}, \bibinfo {author}
  {\bibfnamefont {J.~M.~J.}\ \bibnamefont {Keeling}}, \bibinfo {author}
  {\bibfnamefont {F.~M.}\ \bibnamefont {Marchetti}}, \bibinfo {author}
  {\bibfnamefont {M.~H.}\ \bibnamefont {Szymanska}}, \bibinfo {author}
  {\bibfnamefont {R.}~\bibnamefont {Andre}}, \bibinfo {author} {\bibfnamefont
  {J.~L.}\ \bibnamefont {Staehli}}, \bibinfo {author} {\bibfnamefont
  {V.}~\bibnamefont {Savona}}, \bibinfo {author} {\bibfnamefont {P.~B.}\
  \bibnamefont {Littlewood}}, \bibinfo {author} {\bibfnamefont
  {B.}~\bibnamefont {Deveaud}}, \ and\ \bibinfo {author} {\bibfnamefont
  {L.~S.}\ \bibnamefont {Dang}},\ }\href@noop {} {\bibfield  {journal}
  {\bibinfo  {journal} {Nature}\ }\textbf {\bibinfo {volume} {\textbf{443}}},\
  \bibinfo {pages} {409} (\bibinfo {year} {2006})}\BibitemShut {NoStop}%
\bibitem [{\citenamefont {Amo}\ \emph {et~al.}(2009{\natexlab{a}})\citenamefont
  {Amo}, \citenamefont {Lefr\`{e}re}, \citenamefont {Pigeon}, \citenamefont
  {Adrados}, \citenamefont {Ciuti}, \citenamefont {Carusotto}, \citenamefont
  {Houdr\'{e}}, \citenamefont {Giacobino},\ and\ \citenamefont
  {Bramati}}]{Amo2008}%
  \BibitemOpen
  \bibfield  {author} {\bibinfo {author} {\bibfnamefont {A.}~\bibnamefont
  {Amo}}, \bibinfo {author} {\bibfnamefont {J.}~\bibnamefont {Lefr\`{e}re}},
  \bibinfo {author} {\bibfnamefont {S.}~\bibnamefont {Pigeon}}, \bibinfo
  {author} {\bibfnamefont {C.}~\bibnamefont {Adrados}}, \bibinfo {author}
  {\bibfnamefont {C.}~\bibnamefont {Ciuti}}, \bibinfo {author} {\bibfnamefont
  {I.}~\bibnamefont {Carusotto}}, \bibinfo {author} {\bibfnamefont
  {R.}~\bibnamefont {Houdr\'{e}}}, \bibinfo {author} {\bibfnamefont
  {E.}~\bibnamefont {Giacobino}}, \ and\ \bibinfo {author} {\bibfnamefont
  {A.}~\bibnamefont {Bramati}},\ }\href@noop {} {\bibfield  {journal} {\bibinfo
   {journal} {Nature Phys.}\ }\textbf {\bibinfo {volume} {\textbf{5}}},\
  \bibinfo {pages} {805} (\bibinfo {year} {2009}{\natexlab{a}})}\BibitemShut
  {NoStop}%
\bibitem [{\citenamefont {Lagoudakis}\ \emph {et~al.}(2008)\citenamefont
  {Lagoudakis}, \citenamefont {Wouters}, \citenamefont {Richard}, \citenamefont
  {Baas}, \citenamefont {Carusotto}, \citenamefont {Andre}, \citenamefont
  {Dang},\ and\ \citenamefont {Deveaud-Pledran}}]{Lagoudakis2008}%
  \BibitemOpen
  \bibfield  {author} {\bibinfo {author} {\bibfnamefont {K.~G.}\ \bibnamefont
  {Lagoudakis}}, \bibinfo {author} {\bibfnamefont {M.}~\bibnamefont {Wouters}},
  \bibinfo {author} {\bibfnamefont {M.}~\bibnamefont {Richard}}, \bibinfo
  {author} {\bibfnamefont {A.}~\bibnamefont {Baas}}, \bibinfo {author}
  {\bibfnamefont {I.}~\bibnamefont {Carusotto}}, \bibinfo {author}
  {\bibfnamefont {R.}~\bibnamefont {Andre}}, \bibinfo {author} {\bibfnamefont
  {L.~S.}\ \bibnamefont {Dang}}, \ and\ \bibinfo {author} {\bibfnamefont
  {B.}~\bibnamefont {Deveaud-Pledran}},\ }\href@noop {} {\bibfield  {journal}
  {\bibinfo  {journal} {Nature Phys.}\ }\textbf {\bibinfo {volume}
  {\textbf{4}}},\ \bibinfo {pages} {706} (\bibinfo {year} {2008})}\BibitemShut
  {NoStop}%
\bibitem [{\citenamefont {Amo}\ \emph {et~al.}(2011)\citenamefont {Amo},
  \citenamefont {Pigeon}, \citenamefont {Sanvitto}, \citenamefont {Sala},
  \citenamefont {Hivet}, \citenamefont {Carusotto}, \citenamefont {Pisanello},
  \citenamefont {Lemenager}, \citenamefont {Houdre}, \citenamefont {Giacobino},
  \citenamefont {Ciuti},\ and\ \citenamefont {Bramati}}]{Amo2011}%
  \BibitemOpen
  \bibfield  {author} {\bibinfo {author} {\bibfnamefont {A.}~\bibnamefont
  {Amo}}, \bibinfo {author} {\bibfnamefont {S.}~\bibnamefont {Pigeon}},
  \bibinfo {author} {\bibfnamefont {D.}~\bibnamefont {Sanvitto}}, \bibinfo
  {author} {\bibfnamefont {V.~G.}\ \bibnamefont {Sala}}, \bibinfo {author}
  {\bibfnamefont {R.}~\bibnamefont {Hivet}}, \bibinfo {author} {\bibfnamefont
  {I.}~\bibnamefont {Carusotto}}, \bibinfo {author} {\bibfnamefont
  {F.}~\bibnamefont {Pisanello}}, \bibinfo {author} {\bibfnamefont
  {G.}~\bibnamefont {Lemenager}}, \bibinfo {author} {\bibfnamefont
  {R.}~\bibnamefont {Houdre}}, \bibinfo {author} {\bibfnamefont
  {E.}~\bibnamefont {Giacobino}}, \bibinfo {author} {\bibfnamefont
  {C.}~\bibnamefont {Ciuti}}, \ and\ \bibinfo {author} {\bibfnamefont
  {A.}~\bibnamefont {Bramati}},\ }\href@noop {} {\bibfield  {journal} {\bibinfo
   {journal} {Science}\ }\textbf {\bibinfo {volume} {332}},\ \bibinfo {pages}
  {1167} (\bibinfo {year} {2011})}\BibitemShut {NoStop}%
\bibitem [{\citenamefont {Sich}\ \emph {et~al.}(2012)\citenamefont {Sich},
  \citenamefont {Krizhanovskii}, \citenamefont {Skolnick}, \citenamefont
  {Gorbach}, \citenamefont {Hartley}, \citenamefont {Skryabin}, \citenamefont
  {Cerda-Mendez}, \citenamefont {Biermann}, \citenamefont {Hey},\ and\
  \citenamefont {Santos}}]{Sich2012}%
  \BibitemOpen
  \bibfield  {author} {\bibinfo {author} {\bibfnamefont {M.}~\bibnamefont
  {Sich}}, \bibinfo {author} {\bibfnamefont {D.~N.}\ \bibnamefont
  {Krizhanovskii}}, \bibinfo {author} {\bibfnamefont {M.~S.}\ \bibnamefont
  {Skolnick}}, \bibinfo {author} {\bibfnamefont {A.~V.}\ \bibnamefont
  {Gorbach}}, \bibinfo {author} {\bibfnamefont {R.}~\bibnamefont {Hartley}},
  \bibinfo {author} {\bibfnamefont {D.~V.}\ \bibnamefont {Skryabin}}, \bibinfo
  {author} {\bibfnamefont {E.~A.}\ \bibnamefont {Cerda-Mendez}}, \bibinfo
  {author} {\bibfnamefont {K.}~\bibnamefont {Biermann}}, \bibinfo {author}
  {\bibfnamefont {R.}~\bibnamefont {Hey}}, \ and\ \bibinfo {author}
  {\bibfnamefont {P.~V.}\ \bibnamefont {Santos}},\ }\href@noop {} {\bibfield
  {journal} {\bibinfo  {journal} {Nature Phot.}\ }\textbf {\bibinfo {volume}
  {6}},\ \bibinfo {pages} {50} (\bibinfo {year} {2012})}\BibitemShut {NoStop}%
\bibitem [{\citenamefont {Wertz}\ \emph {et~al.}(2010)\citenamefont {Wertz},
  \citenamefont {Ferrier}, \citenamefont {Solnyshkov}, \citenamefont {Johne},
  \citenamefont {Sanvitto}, \citenamefont {Lemaitre}, \citenamefont {Sagnes},
  \citenamefont {Grousson}, \citenamefont {Kavokin}, \citenamefont {Senellart},
  \citenamefont {Malpuech},\ and\ \citenamefont {Bloch}}]{Wertz2010}%
  \BibitemOpen
  \bibfield  {author} {\bibinfo {author} {\bibfnamefont {E.}~\bibnamefont
  {Wertz}}, \bibinfo {author} {\bibfnamefont {L.}~\bibnamefont {Ferrier}},
  \bibinfo {author} {\bibfnamefont {D.~D.}\ \bibnamefont {Solnyshkov}},
  \bibinfo {author} {\bibfnamefont {R.}~\bibnamefont {Johne}}, \bibinfo
  {author} {\bibfnamefont {D.}~\bibnamefont {Sanvitto}}, \bibinfo {author}
  {\bibfnamefont {A.}~\bibnamefont {Lemaitre}}, \bibinfo {author}
  {\bibfnamefont {I.}~\bibnamefont {Sagnes}}, \bibinfo {author} {\bibfnamefont
  {R.}~\bibnamefont {Grousson}}, \bibinfo {author} {\bibfnamefont {A.~V.}\
  \bibnamefont {Kavokin}}, \bibinfo {author} {\bibfnamefont {P.}~\bibnamefont
  {Senellart}}, \bibinfo {author} {\bibfnamefont {G.}~\bibnamefont {Malpuech}},
  \ and\ \bibinfo {author} {\bibfnamefont {J.}~\bibnamefont {Bloch}},\
  }\href@noop {} {\bibfield  {journal} {\bibinfo  {journal} {Nature Phys.}\
  }\textbf {\bibinfo {volume} {6}},\ \bibinfo {pages} {860} (\bibinfo {year}
  {2010})}\BibitemShut {NoStop}%
\bibitem [{\citenamefont {Christmann}\ \emph {et~al.}(2012)\citenamefont
  {Christmann}, \citenamefont {Tosi}, \citenamefont {Berloff}, \citenamefont
  {Tsotsis}, \citenamefont {Eldridge}, \citenamefont {Hatzopoulos},
  \citenamefont {Savvidis},\ and\ \citenamefont {Baumberg}}]{Christmann2012}%
  \BibitemOpen
  \bibfield  {author} {\bibinfo {author} {\bibfnamefont {G.}~\bibnamefont
  {Christmann}}, \bibinfo {author} {\bibfnamefont {G.}~\bibnamefont {Tosi}},
  \bibinfo {author} {\bibfnamefont {N.~G.}\ \bibnamefont {Berloff}}, \bibinfo
  {author} {\bibfnamefont {P.}~\bibnamefont {Tsotsis}}, \bibinfo {author}
  {\bibfnamefont {P.~S.}\ \bibnamefont {Eldridge}}, \bibinfo {author}
  {\bibfnamefont {Z.}~\bibnamefont {Hatzopoulos}}, \bibinfo {author}
  {\bibfnamefont {P.~G.}\ \bibnamefont {Savvidis}}, \ and\ \bibinfo {author}
  {\bibfnamefont {J.~J.}\ \bibnamefont {Baumberg}},\ }\href@noop {} {\bibfield
  {journal} {\bibinfo  {journal} {Phys. Rev. B}\ }\textbf {\bibinfo {volume}
  {85}},\ \bibinfo {pages} {235303} (\bibinfo {year} {2012})}\BibitemShut
  {NoStop}%
\bibitem [{\citenamefont {Shelykh}\ \emph {et~al.}(2009)\citenamefont
  {Shelykh}, \citenamefont {Pavlovic}, \citenamefont {Solnyshkov},\ and\
  \citenamefont {Malpuech}}]{Shelykh2009}%
  \BibitemOpen
  \bibfield  {author} {\bibinfo {author} {\bibfnamefont {I.~A.}\ \bibnamefont
  {Shelykh}}, \bibinfo {author} {\bibfnamefont {G.}~\bibnamefont {Pavlovic}},
  \bibinfo {author} {\bibfnamefont {D.~D.}\ \bibnamefont {Solnyshkov}}, \ and\
  \bibinfo {author} {\bibfnamefont {G.}~\bibnamefont {Malpuech}},\ }\href@noop
  {} {\bibfield  {journal} {\bibinfo  {journal} {Phys. Rev. Lett.}\ }\textbf
  {\bibinfo {volume} {102}},\ \bibinfo {pages} {046407} (\bibinfo {year}
  {2009})}\BibitemShut {NoStop}%
\bibitem [{\citenamefont {Liew}\ \emph {et~al.}(2008)\citenamefont {Liew},
  \citenamefont {Kavokin},\ and\ \citenamefont {Shelykh}}]{Liew2008}%
  \BibitemOpen
  \bibfield  {author} {\bibinfo {author} {\bibfnamefont {T.~C.~H.}\
  \bibnamefont {Liew}}, \bibinfo {author} {\bibfnamefont {A.~V.}\ \bibnamefont
  {Kavokin}}, \ and\ \bibinfo {author} {\bibfnamefont {I.~A.}\ \bibnamefont
  {Shelykh}},\ }\href@noop {} {\bibfield  {journal} {\bibinfo  {journal} {Phys.
  Rev. Lett.}\ }\textbf {\bibinfo {volume} {101}},\ \bibinfo {pages} {016402}
  (\bibinfo {year} {2008})}\BibitemShut {NoStop}%
\bibitem [{\citenamefont {Johne}\ \emph {et~al.}(2010)\citenamefont {Johne},
  \citenamefont {Shelykh}, \citenamefont {Solnyshkov},\ and\ \citenamefont
  {Malpuech}}]{Johne2010}%
  \BibitemOpen
  \bibfield  {author} {\bibinfo {author} {\bibfnamefont {R.}~\bibnamefont
  {Johne}}, \bibinfo {author} {\bibfnamefont {I.~A.}\ \bibnamefont {Shelykh}},
  \bibinfo {author} {\bibfnamefont {D.~D.}\ \bibnamefont {Solnyshkov}}, \ and\
  \bibinfo {author} {\bibfnamefont {G.}~\bibnamefont {Malpuech}},\ }\href@noop
  {} {\bibfield  {journal} {\bibinfo  {journal} {Phys. Rev. B}\ }\textbf
  {\bibinfo {volume} {81}},\ \bibinfo {pages} {125327} (\bibinfo {year}
  {2010})}\BibitemShut {NoStop}%
\bibitem [{\citenamefont {Amo}\ \emph {et~al.}(2010)\citenamefont {Amo},
  \citenamefont {Liew}, \citenamefont {Adrados}, \citenamefont {Houdre},
  \citenamefont {Giacobino}, \citenamefont {Kavokin},\ and\ \citenamefont
  {Bramati}}]{Amo2010}%
  \BibitemOpen
  \bibfield  {author} {\bibinfo {author} {\bibfnamefont {A.}~\bibnamefont
  {Amo}}, \bibinfo {author} {\bibfnamefont {T.~C.~H.}\ \bibnamefont {Liew}},
  \bibinfo {author} {\bibfnamefont {C.}~\bibnamefont {Adrados}}, \bibinfo
  {author} {\bibfnamefont {R.}~\bibnamefont {Houdre}}, \bibinfo {author}
  {\bibfnamefont {E.}~\bibnamefont {Giacobino}}, \bibinfo {author}
  {\bibfnamefont {A.~V.}\ \bibnamefont {Kavokin}}, \ and\ \bibinfo {author}
  {\bibfnamefont {A.}~\bibnamefont {Bramati}},\ }\href@noop {} {\bibfield
  {journal} {\bibinfo  {journal} {Nature Phot.}\ }\textbf {\bibinfo {volume}
  {4}},\ \bibinfo {pages} {361} (\bibinfo {year} {2010})}\BibitemShut {NoStop}%
\bibitem [{\citenamefont {Shelykh}\ \emph {et~al.}(2010)\citenamefont
  {Shelykh}, \citenamefont {Johne}, \citenamefont {Solnyshkov},\ and\
  \citenamefont {Malpuech}}]{Shelykh2010b}%
  \BibitemOpen
  \bibfield  {author} {\bibinfo {author} {\bibfnamefont {I.~A.}\ \bibnamefont
  {Shelykh}}, \bibinfo {author} {\bibfnamefont {R.}~\bibnamefont {Johne}},
  \bibinfo {author} {\bibfnamefont {D.~D.}\ \bibnamefont {Solnyshkov}}, \ and\
  \bibinfo {author} {\bibfnamefont {G.}~\bibnamefont {Malpuech}},\ }\href@noop
  {} {\bibfield  {journal} {\bibinfo  {journal} {Phys. Rev. B}\ }\textbf
  {\bibinfo {volume} {82}},\ \bibinfo {pages} {153303} (\bibinfo {year}
  {2010})}\BibitemShut {NoStop}%
\bibitem [{\citenamefont {Flayac}\ \emph {et~al.}(2011)\citenamefont {Flayac},
  \citenamefont {Solnyshkov},\ and\ \citenamefont {Malpuech}}]{Flayac2011b}%
  \BibitemOpen
  \bibfield  {author} {\bibinfo {author} {\bibfnamefont {H.}~\bibnamefont
  {Flayac}}, \bibinfo {author} {\bibfnamefont {D.~D.}\ \bibnamefont
  {Solnyshkov}}, \ and\ \bibinfo {author} {\bibfnamefont {G.}~\bibnamefont
  {Malpuech}},\ }\href@noop {} {\bibfield  {journal} {\bibinfo  {journal}
  {Phys. Rev. B}\ }\textbf {\bibinfo {volume} {84}},\ \bibinfo {pages} {125314}
  (\bibinfo {year} {2011})}\BibitemShut {NoStop}%
\bibitem [{\citenamefont {Bloch}\ and\ \citenamefont
  {Marzin}(1997)}]{Bloch1997}%
  \BibitemOpen
  \bibfield  {author} {\bibinfo {author} {\bibfnamefont {J.}~\bibnamefont
  {Bloch}}\ and\ \bibinfo {author} {\bibfnamefont {J.~Y.}\ \bibnamefont
  {Marzin}},\ }\href@noop {} {\bibfield  {journal} {\bibinfo  {journal} {Phys.
  Rev. B}\ }\textbf {\bibinfo {volume} {\textbf{56}}},\ \bibinfo {pages} {2103}
  (\bibinfo {year} {1997})}\BibitemShut {NoStop}%
\bibitem [{\citenamefont {Damen}\ \emph {et~al.}(1990)\citenamefont {Damen},
  \citenamefont {Shah}, \citenamefont {Oberli}, \citenamefont {Chemla},
  \citenamefont {Cunningham},\ and\ \citenamefont {Kuo}}]{Damen1990}%
  \BibitemOpen
  \bibfield  {author} {\bibinfo {author} {\bibfnamefont {T.~C.}\ \bibnamefont
  {Damen}}, \bibinfo {author} {\bibfnamefont {J.}~\bibnamefont {Shah}},
  \bibinfo {author} {\bibfnamefont {D.~Y.}\ \bibnamefont {Oberli}}, \bibinfo
  {author} {\bibfnamefont {D.~S.}\ \bibnamefont {Chemla}}, \bibinfo {author}
  {\bibfnamefont {J.~E.}\ \bibnamefont {Cunningham}}, \ and\ \bibinfo {author}
  {\bibfnamefont {J.~M.}\ \bibnamefont {Kuo}},\ }\href@noop {} {\bibfield
  {journal} {\bibinfo  {journal} {Phys. Rev. B}\ }\textbf {\bibinfo {volume}
  {\textbf{42}}},\ \bibinfo {pages} {7434} (\bibinfo {year}
  {1990})}\BibitemShut {NoStop}%
\bibitem [{\citenamefont {Szczytko}\ \emph {et~al.}(2004)\citenamefont
  {Szczytko}, \citenamefont {Kappei}, \citenamefont {Berney}, \citenamefont
  {Morier-Genoud}, \citenamefont {Portella-Oberli},\ and\ \citenamefont
  {Deveaud}}]{Szczytko2004}%
  \BibitemOpen
  \bibfield  {author} {\bibinfo {author} {\bibfnamefont {J.}~\bibnamefont
  {Szczytko}}, \bibinfo {author} {\bibfnamefont {L.}~\bibnamefont {Kappei}},
  \bibinfo {author} {\bibfnamefont {J.}~\bibnamefont {Berney}}, \bibinfo
  {author} {\bibfnamefont {F.}~\bibnamefont {Morier-Genoud}}, \bibinfo {author}
  {\bibfnamefont {M.}~\bibnamefont {Portella-Oberli}}, \ and\ \bibinfo {author}
  {\bibfnamefont {B.}~\bibnamefont {Deveaud}},\ }\href@noop {} {\bibfield
  {journal} {\bibinfo  {journal} {Phys. Rev. Lett.}\ }\textbf {\bibinfo
  {volume} {\textbf{93}}},\ \bibinfo {pages} {137401} (\bibinfo {year}
  {2004})}\BibitemShut {NoStop}%
\bibitem [{\citenamefont {Ferrier}\ \emph {et~al.}(2011)\citenamefont
  {Ferrier}, \citenamefont {Wertz}, \citenamefont {Johne}, \citenamefont
  {Solnyshkov}, \citenamefont {Senellart}, \citenamefont {Sagnes},
  \citenamefont {Lemaitre}, \citenamefont {Malpuech},\ and\ \citenamefont
  {Bloch}}]{Ferrier2011}%
  \BibitemOpen
  \bibfield  {author} {\bibinfo {author} {\bibfnamefont {L.}~\bibnamefont
  {Ferrier}}, \bibinfo {author} {\bibfnamefont {E.}~\bibnamefont {Wertz}},
  \bibinfo {author} {\bibfnamefont {R.}~\bibnamefont {Johne}}, \bibinfo
  {author} {\bibfnamefont {D.~D.}\ \bibnamefont {Solnyshkov}}, \bibinfo
  {author} {\bibfnamefont {P.}~\bibnamefont {Senellart}}, \bibinfo {author}
  {\bibfnamefont {I.}~\bibnamefont {Sagnes}}, \bibinfo {author} {\bibfnamefont
  {A.}~\bibnamefont {Lemaitre}}, \bibinfo {author} {\bibfnamefont
  {G.}~\bibnamefont {Malpuech}}, \ and\ \bibinfo {author} {\bibfnamefont
  {J.}~\bibnamefont {Bloch}},\ }\href@noop {} {\bibfield  {journal} {\bibinfo
  {journal} {Phys. Rev. Lett.}\ }\textbf {\bibinfo {volume} {106}},\ \bibinfo
  {pages} {126401} (\bibinfo {year} {2011})}\BibitemShut {NoStop}%
\bibitem [{\citenamefont {Tanese}\ \emph {et~al.}(2012)\citenamefont {Tanese},
  \citenamefont {Solnyshkov}, \citenamefont {Amo}, \citenamefont {Ferrier},
  \citenamefont {Bernet-Rollande}, \citenamefont {Wertz}, \citenamefont
  {Sagnes}, \citenamefont {Lema\^itre}, \citenamefont {Senellart},
  \citenamefont {Malpuech},\ and\ \citenamefont {Bloch}}]{Tanese2012}%
  \BibitemOpen
  \bibfield  {author} {\bibinfo {author} {\bibfnamefont {D.}~\bibnamefont
  {Tanese}}, \bibinfo {author} {\bibfnamefont {D.~D.}\ \bibnamefont
  {Solnyshkov}}, \bibinfo {author} {\bibfnamefont {A.}~\bibnamefont {Amo}},
  \bibinfo {author} {\bibfnamefont {L.}~\bibnamefont {Ferrier}}, \bibinfo
  {author} {\bibfnamefont {E.}~\bibnamefont {Bernet-Rollande}}, \bibinfo
  {author} {\bibfnamefont {E.}~\bibnamefont {Wertz}}, \bibinfo {author}
  {\bibfnamefont {I.}~\bibnamefont {Sagnes}}, \bibinfo {author} {\bibfnamefont
  {A.}~\bibnamefont {Lema\^itre}}, \bibinfo {author} {\bibfnamefont
  {P.}~\bibnamefont {Senellart}}, \bibinfo {author} {\bibfnamefont
  {G.}~\bibnamefont {Malpuech}}, \ and\ \bibinfo {author} {\bibfnamefont
  {J.}~\bibnamefont {Bloch}},\ }\href@noop {} {\bibfield  {journal} {\bibinfo
  {journal} {Phys. Rev. Lett.}\ }\textbf {\bibinfo {volume} {108}},\ \bibinfo
  {pages} {036405} (\bibinfo {year} {2012})}\BibitemShut {NoStop}%
\bibitem [{Note1()}]{Note1}%
  \BibitemOpen
  \bibinfo {note} {See Supplementary Information for a complete description of
  the model and additional results under high density excitation}\BibitemShut
  {NoStop}%
\bibitem [{\citenamefont {Freixanet}\ \emph {et~al.}(2000)\citenamefont
  {Freixanet}, \citenamefont {Sermage}, \citenamefont {Tiberj},\ and\
  \citenamefont {Planel}}]{Freixanet2000}%
  \BibitemOpen
  \bibfield  {author} {\bibinfo {author} {\bibfnamefont {T.}~\bibnamefont
  {Freixanet}}, \bibinfo {author} {\bibfnamefont {B.}~\bibnamefont {Sermage}},
  \bibinfo {author} {\bibfnamefont {A.}~\bibnamefont {Tiberj}}, \ and\ \bibinfo
  {author} {\bibfnamefont {R.}~\bibnamefont {Planel}},\ }\href@noop {}
  {\bibfield  {journal} {\bibinfo  {journal} {Phys. Rev. B}\ }\textbf {\bibinfo
  {volume} {\textbf{61}}},\ \bibinfo {pages} {7233} (\bibinfo {year}
  {2000})}\BibitemShut {NoStop}%
\bibitem [{\citenamefont {Choi}\ \emph {et~al.}(1998)\citenamefont {Choi},
  \citenamefont {Morgan},\ and\ \citenamefont {Burnett}}]{Choi1998}%
  \BibitemOpen
  \bibfield  {author} {\bibinfo {author} {\bibfnamefont {S.}~\bibnamefont
  {Choi}}, \bibinfo {author} {\bibfnamefont {S.~A.}\ \bibnamefont {Morgan}}, \
  and\ \bibinfo {author} {\bibfnamefont {K.}~\bibnamefont {Burnett}},\ }\href
  {\doibase 10.1103/PhysRevA.57.4057} {\bibfield  {journal} {\bibinfo
  {journal} {Phys. Rev. A}\ }\textbf {\bibinfo {volume} {57}},\ \bibinfo
  {pages} {4057} (\bibinfo {year} {1998})}\BibitemShut {NoStop}%
\bibitem [{\citenamefont {Wouters}\ \emph {et~al.}(2010)\citenamefont
  {Wouters}, \citenamefont {Liew},\ and\ \citenamefont
  {Savona}}]{Wouters2010c}%
  \BibitemOpen
  \bibfield  {author} {\bibinfo {author} {\bibfnamefont {M.}~\bibnamefont
  {Wouters}}, \bibinfo {author} {\bibfnamefont {T.~C.~H.}\ \bibnamefont
  {Liew}}, \ and\ \bibinfo {author} {\bibfnamefont {V.}~\bibnamefont
  {Savona}},\ }\href@noop {} {\bibfield  {journal} {\bibinfo  {journal} {Phys.
  Rev. B}\ }\textbf {\bibinfo {volume} {82}},\ \bibinfo {pages} {245315}
  (\bibinfo {year} {2010})}\BibitemShut {NoStop}%
\bibitem [{\citenamefont {Wouters}\ and\ \citenamefont
  {Savona}(2010)}]{Wouters2010d}%
  \BibitemOpen
  \bibfield  {author} {\bibinfo {author} {\bibfnamefont {M.}~\bibnamefont
  {Wouters}}\ and\ \bibinfo {author} {\bibfnamefont {V.}~\bibnamefont
  {Savona}},\ }\href@noop {} {\bibfield  {journal} {\bibinfo  {journal}
  {arXiv:1007.5453v1}\ } (\bibinfo {year} {2010})}\BibitemShut {NoStop}%
\bibitem [{\citenamefont {Gao}\ \emph {et~al.}(2012)\citenamefont {Gao},
  \citenamefont {Eldridge}, \citenamefont {Liew}, \citenamefont {Tsintzos},
  \citenamefont {Stavrinidis}, \citenamefont {Deligeorgis}, \citenamefont
  {Hatzopoulos},\ and\ \citenamefont {Savvidis}}]{Gao2012}%
  \BibitemOpen
  \bibfield  {author} {\bibinfo {author} {\bibfnamefont {T.}~\bibnamefont
  {Gao}}, \bibinfo {author} {\bibfnamefont {P.~S.}\ \bibnamefont {Eldridge}},
  \bibinfo {author} {\bibfnamefont {T.~C.~H.}\ \bibnamefont {Liew}}, \bibinfo
  {author} {\bibfnamefont {S.~I.}\ \bibnamefont {Tsintzos}}, \bibinfo {author}
  {\bibfnamefont {G.}~\bibnamefont {Stavrinidis}}, \bibinfo {author}
  {\bibfnamefont {G.}~\bibnamefont {Deligeorgis}}, \bibinfo {author}
  {\bibfnamefont {Z.}~\bibnamefont {Hatzopoulos}}, \ and\ \bibinfo {author}
  {\bibfnamefont {P.~G.}\ \bibnamefont {Savvidis}},\ }\href@noop {} {\bibfield
  {journal} {\bibinfo  {journal} {Phys. Rev. B}\ }\textbf {\bibinfo {volume}
  {85}},\ \bibinfo {pages} {235102} (\bibinfo {year} {2012})}\BibitemShut
  {NoStop}%
\bibitem [{\citenamefont {Tosi}\ \emph {et~al.}(2012)\citenamefont {Tosi},
  \citenamefont {Christmann}, \citenamefont {Berloff}, \citenamefont {Tsotsis},
  \citenamefont {Gao}, \citenamefont {Hatzopoulos}, \citenamefont {Savvidis},\
  and\ \citenamefont {Baumberg}}]{Tosi2012}%
  \BibitemOpen
  \bibfield  {author} {\bibinfo {author} {\bibfnamefont {G.}~\bibnamefont
  {Tosi}}, \bibinfo {author} {\bibfnamefont {G.}~\bibnamefont {Christmann}},
  \bibinfo {author} {\bibfnamefont {N.~G.}\ \bibnamefont {Berloff}}, \bibinfo
  {author} {\bibfnamefont {P.}~\bibnamefont {Tsotsis}}, \bibinfo {author}
  {\bibfnamefont {T.}~\bibnamefont {Gao}}, \bibinfo {author} {\bibfnamefont
  {Z.}~\bibnamefont {Hatzopoulos}}, \bibinfo {author} {\bibfnamefont {P.~G.}\
  \bibnamefont {Savvidis}}, \ and\ \bibinfo {author} {\bibfnamefont {J.~J.}\
  \bibnamefont {Baumberg}},\ }\href@noop {} {\bibfield  {journal} {\bibinfo
  {journal} {Nature Phys.}\ }\textbf {\bibinfo {volume} {8}},\ \bibinfo {pages}
  {190} (\bibinfo {year} {2012})}\BibitemShut {NoStop}%
\bibitem [{\citenamefont {Ballarini}\ \emph {et~al.}(2012)\citenamefont
  {Ballarini}, \citenamefont {Giorgi}, \citenamefont {Cancellieri},
  \citenamefont {Houdré}, \citenamefont {Giacobino}, \citenamefont
  {Cingolani}, \citenamefont {Bramati}, \citenamefont {Gigli},\ and\
  \citenamefont {Sanvitto}}]{Ballarini2012}%
  \BibitemOpen
  \bibfield  {author} {\bibinfo {author} {\bibfnamefont {D.}~\bibnamefont
  {Ballarini}}, \bibinfo {author} {\bibfnamefont {M.~D.}\ \bibnamefont
  {Giorgi}}, \bibinfo {author} {\bibfnamefont {E.}~\bibnamefont {Cancellieri}},
  \bibinfo {author} {\bibfnamefont {R.}~\bibnamefont {Houdré}}, \bibinfo
  {author} {\bibfnamefont {E.}~\bibnamefont {Giacobino}}, \bibinfo {author}
  {\bibfnamefont {R.}~\bibnamefont {Cingolani}}, \bibinfo {author}
  {\bibfnamefont {A.}~\bibnamefont {Bramati}}, \bibinfo {author} {\bibfnamefont
  {G.}~\bibnamefont {Gigli}}, \ and\ \bibinfo {author} {\bibfnamefont
  {D.}~\bibnamefont {Sanvitto}},\ }\href@noop {} {\bibfield  {journal}
  {\bibinfo  {journal} {arXiv:1201.4071v1}\ } (\bibinfo {year}
  {2012})}\BibitemShut {NoStop}%
\bibitem [{\citenamefont {Amo}\ \emph {et~al.}(2009{\natexlab{b}})\citenamefont
  {Amo}, \citenamefont {Sanvitto}, \citenamefont {Laussy}, \citenamefont
  {Ballarini}, \citenamefont {del Valle}, \citenamefont {Mart\'in},
  \citenamefont {Lemaitre}, \citenamefont {Bloch}, \citenamefont
  {Krizhanovskii}, \citenamefont {Skolnick}, \citenamefont {Tejedor},\ and\
  \citenamefont {Vi\~na}}]{Amo2009}%
  \BibitemOpen
  \bibfield  {author} {\bibinfo {author} {\bibfnamefont {A.}~\bibnamefont
  {Amo}}, \bibinfo {author} {\bibfnamefont {D.}~\bibnamefont {Sanvitto}},
  \bibinfo {author} {\bibfnamefont {F.~P.}\ \bibnamefont {Laussy}}, \bibinfo
  {author} {\bibfnamefont {D.}~\bibnamefont {Ballarini}}, \bibinfo {author}
  {\bibfnamefont {E.}~\bibnamefont {del Valle}}, \bibinfo {author}
  {\bibfnamefont {M.~D.}\ \bibnamefont {Mart\'in}}, \bibinfo {author}
  {\bibfnamefont {A.}~\bibnamefont {Lemaitre}}, \bibinfo {author}
  {\bibfnamefont {J.}~\bibnamefont {Bloch}}, \bibinfo {author} {\bibfnamefont
  {D.~N.}\ \bibnamefont {Krizhanovskii}}, \bibinfo {author} {\bibfnamefont
  {M.~S.}\ \bibnamefont {Skolnick}}, \bibinfo {author} {\bibfnamefont
  {C.}~\bibnamefont {Tejedor}}, \ and\ \bibinfo {author} {\bibfnamefont
  {L.}~\bibnamefont {Vi\~na}},\ }\href@noop {} {\bibfield  {journal} {\bibinfo
  {journal} {Nature}\ }\textbf {\bibinfo {volume} {\textbf{457}}},\ \bibinfo
  {pages} {291} (\bibinfo {year} {2009}{\natexlab{b}})}\BibitemShut {NoStop}%
\bibitem [{\citenamefont {Adrados}\ \emph {et~al.}(2011)\citenamefont
  {Adrados}, \citenamefont {Liew}, \citenamefont {Amo}, \citenamefont
  {Mart\'in}, \citenamefont {Sanvitto}, \citenamefont {Ant\'on}, \citenamefont
  {Giacobino}, \citenamefont {Kavokin}, \citenamefont {Bramati},\ and\
  \citenamefont {Vi\~na}}]{Adrados2011}%
  \BibitemOpen
  \bibfield  {author} {\bibinfo {author} {\bibfnamefont {C.}~\bibnamefont
  {Adrados}}, \bibinfo {author} {\bibfnamefont {T.~C.~H.}\ \bibnamefont
  {Liew}}, \bibinfo {author} {\bibfnamefont {A.}~\bibnamefont {Amo}}, \bibinfo
  {author} {\bibfnamefont {M.~D.}\ \bibnamefont {Mart\'in}}, \bibinfo {author}
  {\bibfnamefont {D.}~\bibnamefont {Sanvitto}}, \bibinfo {author}
  {\bibfnamefont {C.}~\bibnamefont {Ant\'on}}, \bibinfo {author} {\bibfnamefont
  {E.}~\bibnamefont {Giacobino}}, \bibinfo {author} {\bibfnamefont
  {A.}~\bibnamefont {Kavokin}}, \bibinfo {author} {\bibfnamefont
  {A.}~\bibnamefont {Bramati}}, \ and\ \bibinfo {author} {\bibfnamefont
  {L.}~\bibnamefont {Vi\~na}},\ }\href@noop {} {\bibfield  {journal} {\bibinfo
  {journal} {Phys. Rev. Lett.}\ }\textbf {\bibinfo {volume} {107}},\ \bibinfo
  {pages} {146402} (\bibinfo {year} {2011})}\BibitemShut {NoStop}%
 \bibitem [{\citenamefont {Galbiati}\ \emph {et~al.}(2012)\citenamefont
  {Galbiati}, \citenamefont {Ferrier}, \citenamefont {Solnyshkov},
  \citenamefont {Tanese}, \citenamefont {Wertz}, \citenamefont {Amo},
  \citenamefont {Abbarchi}, \citenamefont {Senellart}, \citenamefont {Sagnes},
  \citenamefont {Lema\^itre}, \citenamefont {Galopin}, \citenamefont
  {Malpuech},\ and\ \citenamefont {Bloch}}]{Galbiati2012}%
  \BibitemOpen
  \bibfield  {author} {\bibinfo {author} {\bibfnamefont {M.}~\bibnamefont
  {Galbiati}}, \bibinfo {author} {\bibfnamefont {L.}~\bibnamefont {Ferrier}},
  \bibinfo {author} {\bibfnamefont {D.~D.}\ \bibnamefont {Solnyshkov}},
  \bibinfo {author} {\bibfnamefont {D.}~\bibnamefont {Tanese}}, \bibinfo
  {author} {\bibfnamefont {E.}~\bibnamefont {Wertz}}, \bibinfo {author}
  {\bibfnamefont {A.}~\bibnamefont {Amo}}, \bibinfo {author} {\bibfnamefont
  {M.}~\bibnamefont {Abbarchi}}, \bibinfo {author} {\bibfnamefont
  {P.}~\bibnamefont {Senellart}}, \bibinfo {author} {\bibfnamefont
  {I.}~\bibnamefont {Sagnes}}, \bibinfo {author} {\bibfnamefont
  {A.}~\bibnamefont {Lema\^itre}}, \bibinfo {author} {\bibfnamefont
  {E.}~\bibnamefont {Galopin}}, \bibinfo {author} {\bibfnamefont
  {G.}~\bibnamefont {Malpuech}}, \ and\ \bibinfo {author} {\bibfnamefont
  {J.}~\bibnamefont {Bloch}},\ }\href {\doibase 10.1103/PhysRevLett.108.126403}
  {\bibfield  {journal} {\bibinfo  {journal} {Phys. Rev. Lett.}\ }\textbf
  {\bibinfo {volume} {108}},\ \bibinfo {pages} {126403} (\bibinfo {year}
  {2012})}\BibitemShut {NoStop}%
\end{thebibliography}
\end{document}